\documentclass[12pt]{article}
\usepackage{epsfig,lineno}

\def\beq{\begin{equation}}
\def\eeq#1{\label{#1}\end{equation}}
\def\eeqn{\end{equation}}

\def\beqa{\begin{eqnarray}}
\def\eeqa#1{\label{#1}\end{eqnarray}}
\def\eeqan{\end{eqnarray}}

\let\bar=\overbar

\def\etal{{\it et al.}}

\def\Dslash{\not{\hbox{\kern-4pt $D$}}}
\def\dslash{\not{\hbox{\kern-2pt $\del$}}}

\def\BR{\mbox{\rm BR}}

\def\msb{{\bar{\ssstyle M \kern -1pt S}}}

\def\s#1{\widetilde{#1}}

\input plb_symbols
\def\lhratio {{\ensuremath{\cal R}}\xspace}

\def\etal {{\em et al.}}

\newcommand{\nbb}        {\mbox{$471\pm3$}}

\def\Bmeson  {\B\ meson}

\def\nonPiee {\ensuremath{123}}

\def\nyPieeRound {\ensuremath{0.6^{+2.5}_{-2.7}}} 
\def\sigPieeRound {\ensuremath{0.4}} 
\def\bfPiee {\ensuremath{0.27^{+1.1}_{-1.2}\pm0.1}} 
\def\ulPiee {\ensuremath{2.3}} 

\def\nonKee {\ensuremath{42}}

\def\nyKeeRound {\ensuremath{0.7^{+1.8}_{-1.2}}}
\def\sigKeeRound {\ensuremath{0.5}} 
\def\bfKee {\ensuremath{0.49^{+1.3}_{-0.8}\pm0.1}} 
\def\ulKee {\ensuremath{3.0}}  

\def\nonPimm {\ensuremath{228}}

\def\nyPimmRound {\ensuremath{0.0^{+3.2}_{-2.0}}}
\def\sigPimmRound {\ensuremath{0.0}} 
\def\bfPimm {\ensuremath{0.03^{+5.1}_{-3.2}\pm0.6}} 
\def\ulPimm {\ensuremath{10.7}}  

\def\nonKmm {\ensuremath{209}}

\def\nyKmmRound {\ensuremath{0.5^{+3.5}_{-2.5}}}
\def\sigKmmRound {\ensuremath{0.2}} 
\def\bfKmm {\ensuremath{0.45^{+3.2}_{-2.7}\pm0.4}} 
\def\ulKmm {\ensuremath{6.7}}  

\def\Title#1{\begin{center} {\Large {\bf #1} } \end{center}}

\begin{document}

\Title{Searches for lepton-number-violating \B\ decays at CLEO, \babar, and \belle}

\bigskip\bigskip


\begin{raggedright}  

{\it Fergus Wilson\index{Wilson, F.F.}\footnote{On behalf of the
    \belle\ and \babar\ collaborations}\\
Rutherford Appleton Laboratory,\\
Chilton, Didcot,\\
Oxon, OX11 0QX, UNITED KINGDOM }
\bigskip\bigskip \\
Proceedings of CKM 2012, the 7th International Workshop on the CKM Unitarity Triangle, University of Cincinnati, USA, 28 September - 2 October 2012 
\end{raggedright}

\section{Introduction}

CLEO, \babar, and \belle\ have searched for lepton-number-violation
(LNV) in \Bmeson\ decays at \epem\ colliders. In the Standard Model (SM), lepton
number $L$ is conserved in low-energy collisions and decays, and the
lepton flavor numbers for the three lepton families are conserved if
neutrinos are massless. A number of mechanisms for LNV have been
proposed including multi-Higgs-boson extensions, leptoquarks and
Majorana neutrinos. 

A common theme among the experiments is the use of two discriminating
variables calculated in the center-of-mass frame (CM) ($\DeltaE= E_B^*
- \sqrt{s}/2$ and $\mes=m_{\rm cand}=\mbc=\sqrt{s/4-p^{*2}_B}$, where
$E_B^*$ and $p_B^*$ are the CM energy and momentum of the
reconstructed \Bmeson\ candidate), and event-shape discriminants. The
signal events peak at zero for \DeltaE, at approximately the \Bmeson\
mass for \mes, and the event-shape is spherical; for the backgrounds,
\DeltaE\ and \mes are smoothly varying, while the event-shape is more
jet-like. The major backgrounds are continuum production of quark
pairs $e^+e^-\to\qqbar$ ($q = u,d,s$ and $c$), semileptonic \B\ decays
which produce a lepton, and other \BB\ decays where a hadron is
mis-identified as a lepton.

\section{Recent results from CLEO, \babar, and \belle}

CLEO reused the techniques for their search for $\B\to X_s
\ellp\ellm$ to look for $\Bp\to h^-\ep\ep$, $h^-\ep\mup$ and
$h^-\mup\mup$ where $h^-= \Km,\Kstarm,\pim$ and
$\rhom$~\cite{bib:cleo}, based on 9.6 million \BB\ events. 
There are three main sources of background: \B\ decays of the type
$\B\to X\jpsi$ and $\B\to X\psitwos$; other \B\ decays, with
two apparent leptons (either real leptons or hadrons misidentified as
leptons); and continuum processes with two apparent leptons.  The
backgrounds from \jpsi\ and \psitwos\ were severe in the searches for
$\B\to X\ellp\ellm$ as the decays provided two opposite sign leptons
but here they contribute less than 0.1 event per decay mode after
careful lepton identification.

An unbinned maximum likelihood (ML) method is used to discriminate
between signal events and the remaining two background sources.  The
variables are: the missing event energy, $\Emiss \equiv 2E_{\rm beam}
- \sum E_{\rm det}$, where $\sum E_{\rm det}$ denotes the sum of
energies of all the detected particles in the event; a Fisher
discriminant ${\cal F}$ based on the event-shape; $M_{\rm cand}$; and
$\DeltaE$.  Loose cuts of $|\Emiss|<2$\gev, $5.2<M_{\rm
cand}<5.3$\gevcc and $|\DeltaE|<0.25$\gev\ are applied. The branching
fraction for the signal and the yields for the two backgrounds are
free parameters in the ML.  No evidence for the decays is found and
all modes have a statistical significance of less than 1.2 standard
deviations. 90\% confidence level (CL) upper limits (UL) are placed on
the branching fractions as shown in Table~\ref{tab:cleo_results}. The
upper limits range from $1.0$ to $8.3\times 10^{-6}$.

\begin{table}[htb]
\begin{center}
\begin{tabular}{lcc lcc lcc}
\hline  
Mode & $S$ & UL &
Mode & $S$ & UL &
Mode & $S$ & UL \\
\hline  
\Km\ep\ep & 0.6$\sigma$ & 1.0 &
\Km\ep\mup & 0.0$\sigma$ & 2.0 &
\Km\mup\mup & 0.0$\sigma$ & 1.8 \\
\Kstarm\ep\ep & 0.0$\sigma$ & 2.8 &
\Kstarm\ep\mup & 0.0$\sigma$ & 4.4 &
\Kstarm\mup\mup & 0.5$\sigma$ & 8.3 \\
\pim\ep\ep & 0.0$\sigma$ & 1.6 &
\pim\ep\mup & 0.0$\sigma$ & 1.3 &
\pim\mup\mup & 0.0$\sigma$ & 1.4 \\
\rhom\ep\ep & 1.1$\sigma$ & 2.6 &
\rhom\ep\mup & 0.3$\sigma$ & 2.2 &
\rhom\mup\mup & 1.0$\sigma$ & 5.0 \\
\hline
\end{tabular}
\caption{CLEO results with the statistical significance $S$ and 90\%
  confidence level upper limits UL ($\times 10^{-6}$) on the branching
  fraction (including systematic error)~\cite{bib:cleo}.}
\label{tab:cleo_results}
\end{center}
\end{table}


\babar\ has repeated the CLEO search for the four modes $\Bp\to h^-
\mup\mup$ and $\Bp\to h^- e^+e^+$, where $h^-=\Km$ or
\pim~\cite{bib:babar1} and they also reuse the techniques developed
for their $\B\to K^{(*)}\ellp\ellm$ analyses, using a data sample of
\nbb\ million \BB\ pairs. The key difference with CLEO is the ML
 function used.  The two main backgrounds from continuum
events and \BB\ decays are suppressed through the use of a ML
 ratio \lhratio\ constructed from boosted decision tree
discriminants (BDTs). Before the ML fit, a selection on \lhratio\
retains 85\% of the simulated signal events while rejecting more than
95\% of the background. The selection efficiency for
simulated signal is 13\%-48\%.

The signal branching fraction and background yields are extracted from
the data with an unbinned ML using \mes\ and
\lhratio. The signal \mes\ distributions are taken from data using a
Gaussian shape unique to each final state, with the mean and width
determined from fits to the analogous final states in the
$\Bp\to\jpsi(\to \ellp\ellm) h^+$ events from the data. No significant
yields are observed and the results of the ML fits to the data are
summarized in Table~\ref{tab:results}.

\begin{table*}[htbp!]
\centering
\begin{tabular}{lrrrrrr}
\hline
Mode & Events & Yield & \calS ($\sigma$) & {\calB}
($\times 10^{-8}$) & $\calB_{UL}$  ($\times 10^{-8}$) \\
\hline
$\Bp\to\pim\ep\ep$   & \nonPiee &
\hspace{1mm} \nyPieeRound & \hspace{1mm} \sigPieeRound & \hspace{1mm} \bfPiee  & \ulPiee  \\
$\Bp\to\Km\ep\ep$    & \nonKee  &  \nyKeeRound & \sigKeeRound & \bfKee  & \ulKee  \\
$\Bp\to\pim\mup\mup$ & \nonPimm &  \nyPimmRound & \sigPimmRound & \bfPimm  & \ulPimm  \\
$\Bp\to\Km\mup\mup$  & \nonKmm  & \nyKmmRound & \sigKmmRound & \bfKmm  & \ulKmm  \\
\hline
\end{tabular}
\caption{\babar\ results, showing the total events in the sample,
  signal yield and its statistical uncertainty, significance \calS,
  branching fraction \BR, and 90\% CL branching fraction upper limit
  $\calB_{UL}$~\cite{bib:babar1}.}
\label{tab:results}
\end{table*}


\belle\ have performed the first searches for the decays $\Bp \to
\Dm\ep\ep$, $\Dm\ep\mup$ and $\Dm\mup\mup$ using a data sample
containing 772 million \BB\ pairs~\cite{bib:belle}. As the CKM matrix
element $\Vcb > \Vub$, these decays could be enhanced by an order of
magnitude compared to $\Bp\to h^-\ellp\ellp$. \belle\ look for an
energetic same-sign dilepton and combine it with a $D$ candidate
requiring a proper charge combination for the dilepton.  
The same-sign lepton
pair must have a total energy in the $\Upsilon(4S)$ center-of-mass
(CM) frame greater than 1.3\gev.  More than 95\% of events have only
one same-sign lepton pair.

Candidate $D^-$ mesons are reconstructed in the $D^-\to K^+\pi^-\pi^-$
decay.  The three tracks from the $D^-$ candidate are fit to a common
vertex and are required to have a $K^+\pi^-\pi^-$ invariant mass
($M_{K\pi\pi}$) within approximately $\pm 10 ~\mevcc$ from the nominal
$D^-$ mass.  The $M_{K\pi\pi}$ distribution is fit to two Gaussian
functions with a common mean in a mass window $\pm 3$ times the width
of the narrower Gaussian component.  The average multiplicity of $D^-$
candidates is 1.3 per event.

The continuum background is rejected using a Fisher discriminant
${\cal F}$ based on event-shape variables, and $\cos\theta_B$, the
cosine of the polar angle of the \B\ candidate flight direction in the
CM frame. Semileptonic decays such as $B \to D^- \ell^+ \nu_\ell X$
have missing energy due to the undetected neutrino and the two
reconstructed leptons do not come from the same vertex. These events
can be rejected using the missing energy $\Emiss$ and $\delta z$, the
separation between the impact parameter of the two leptons in the beam
direction. The four variables, ${\cal F}$, $\cos\theta_B$, $\Emiss$
and $\delta z$, are combined together into a single likelihood ratio,
$R_{\rm s}$. The optimal requirement on $R_{\rm s}$ is determined by
maximizing the figure of merit, $\epsilon_{\rm s}/(a/2 + \sqrt{N_{\rm
b}})$, where $\epsilon_{\rm s}$ is the MC signal efficiency, $N_{\rm
b}$ is the number of expected background events in the signal region,
and $a$ is set to zero.

\begin{figure}[hbt]
  \begin{center}
  \begin{tabular}{ccc}
    \includegraphics[width=0.3\textwidth]{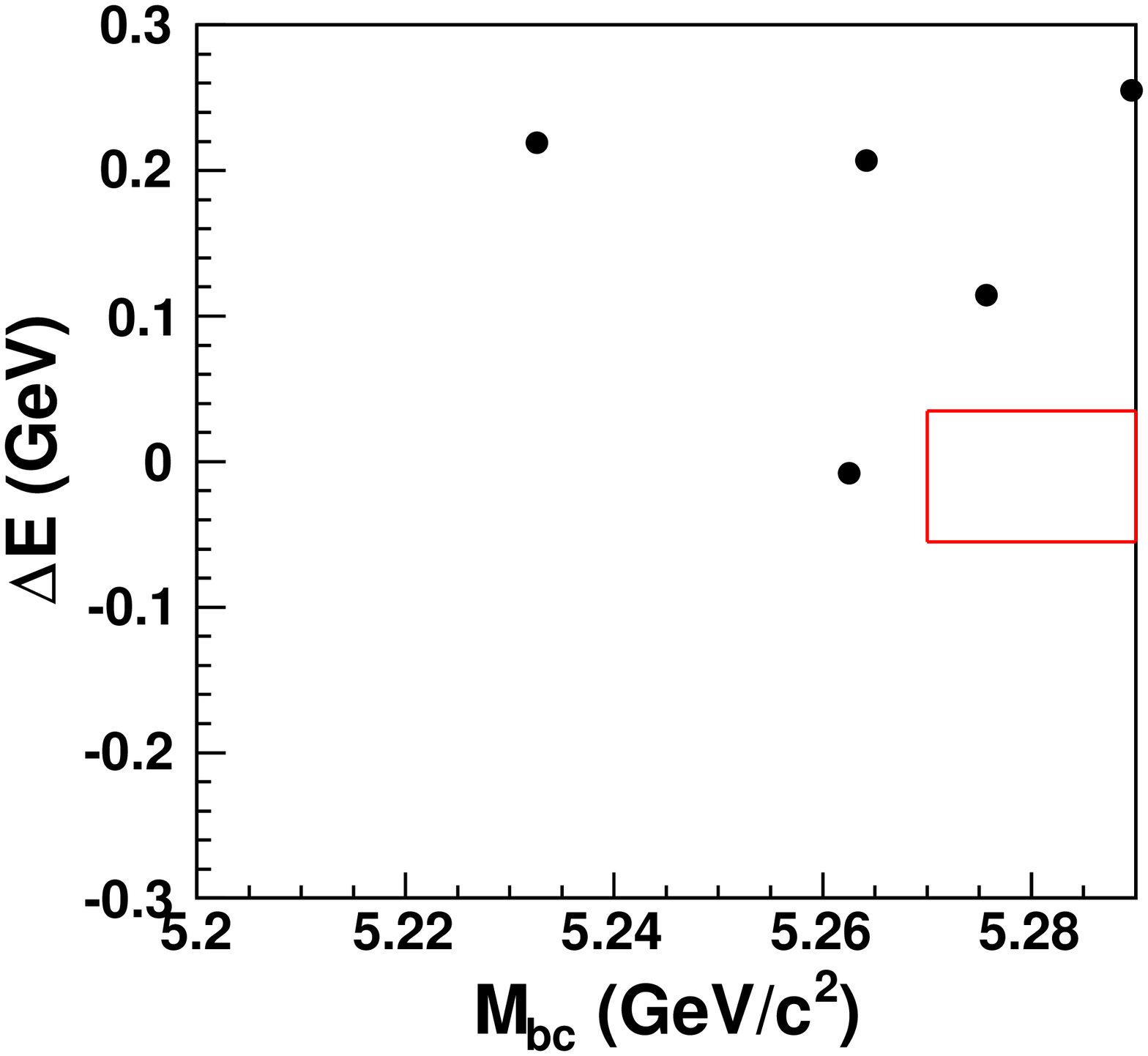} &
    \includegraphics[width=0.3\textwidth]{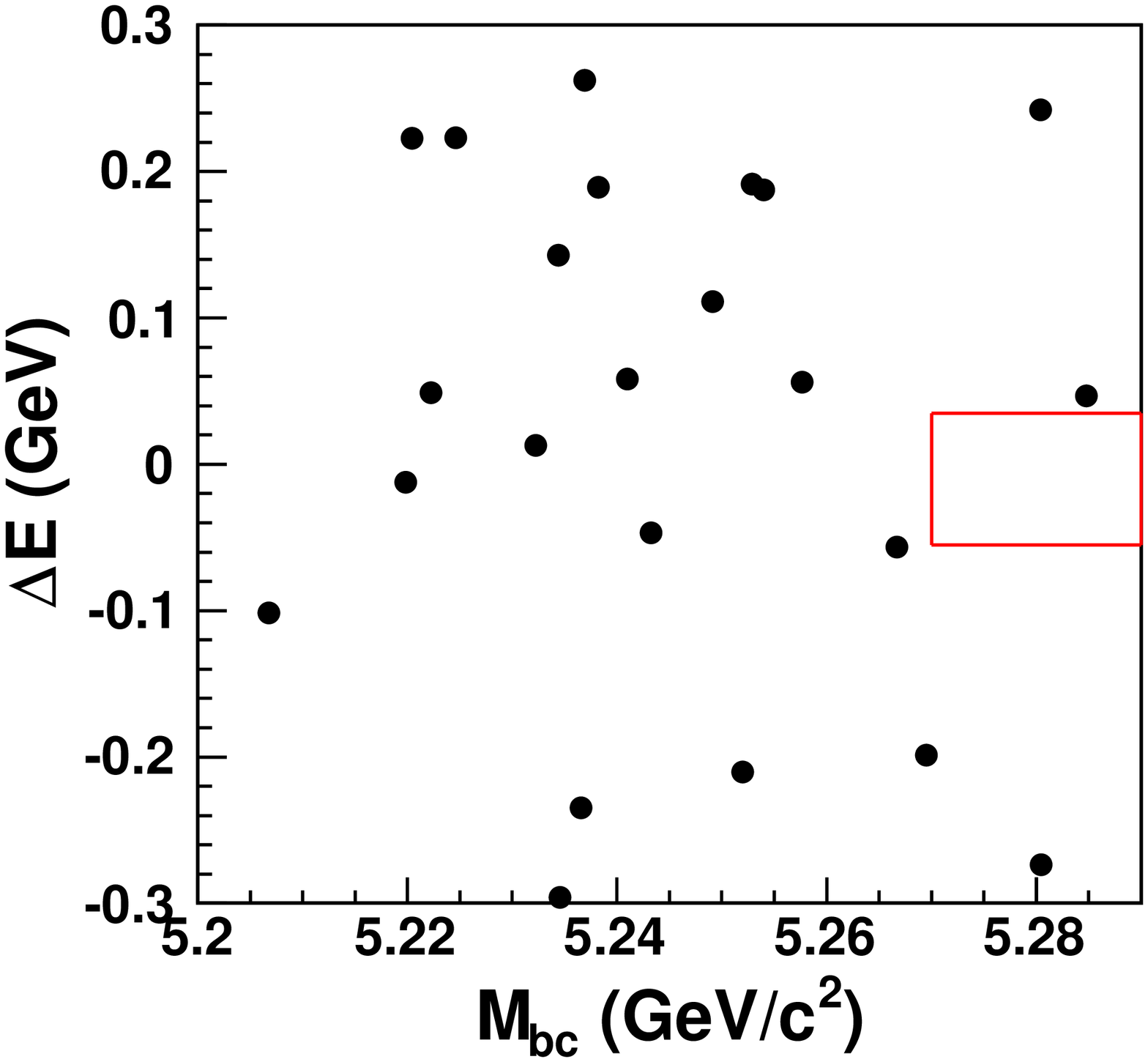} &
    \includegraphics[width=0.3\textwidth]{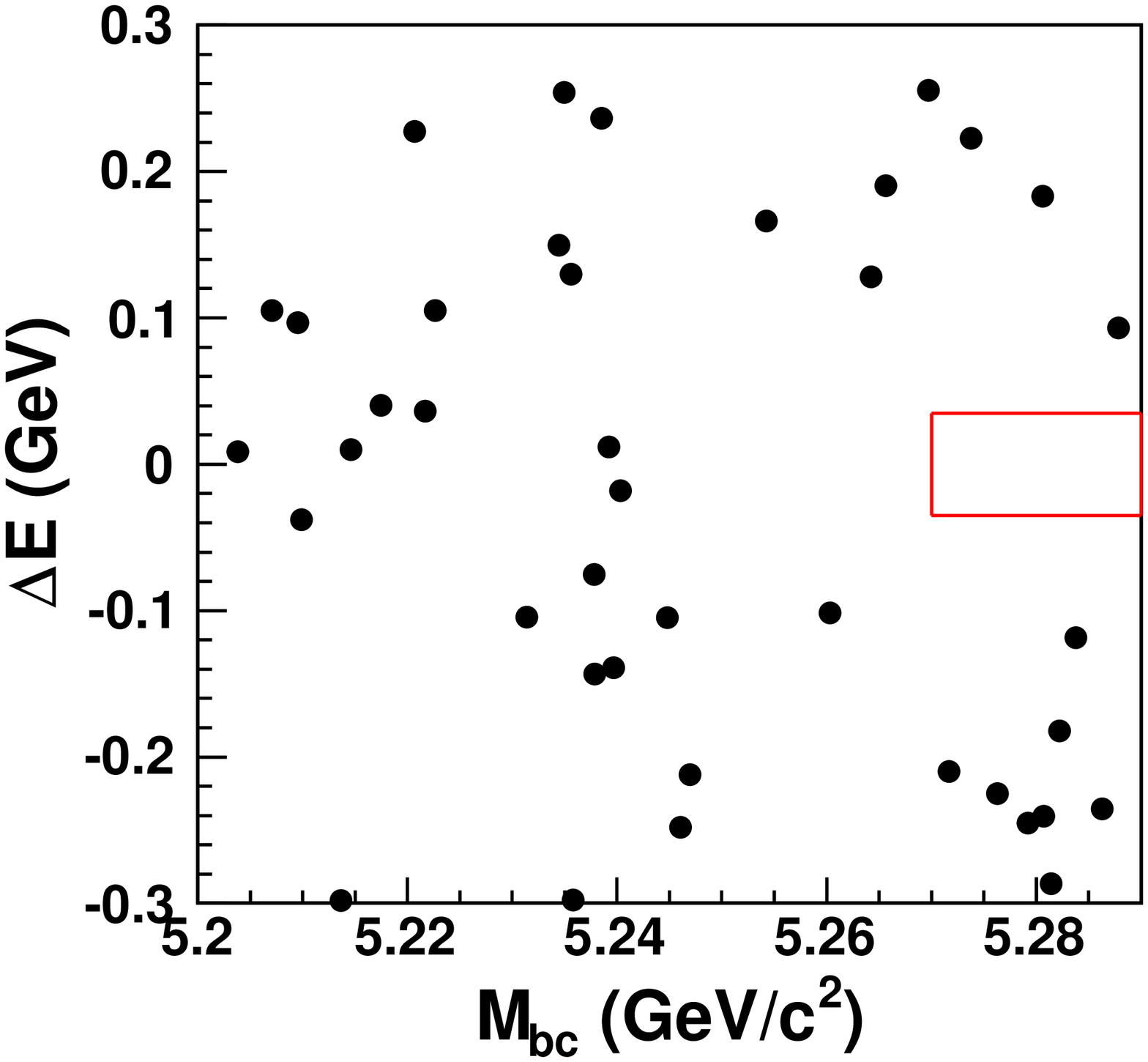}
  \end{tabular}
  \caption{The \mbc-\DeltaE distributions of
    $\Dm\ep\ep$ (left), $\Dm\ep\mup$ (middle) and $\Dm\mup\mup$
    (right) final states in data.  The (red) boxes indicate the signal
    regions.}
  \label{fig:belle_figResults}
  \end{center}
\end{figure}

After applying the $R_{\rm s}$ requirements, 5, 23 and 40 events
remain in the background region for the $e^+e^+$, $e^+\mu^+$ and
$\mu^+\mu^+$ modes, respectively.  The signal efficiencies are
evaluated to be 1.2\% - 1.9\%. The expected numbers of background
events in the signal region are 0.18, 0.83 and 1.44 events for the
$e^+e^+$, $e^+\mu^+$ and $\mu^+\mu^+$ modes,
respectively. Figure~\ref{fig:belle_figResults} shows the
$\mbc$-$\DeltaE$ distributions of the selected events. No events are
observed in the signal region. Table~\ref{tab:belle_result} summarises
the upper limits achieved.

\begin{table}[hbt]
\centering
  \begin{tabular}{ccccc}
  \hline  
  Mode & $\epsilon$ [\%] & $N_{\rm obs}$ & $N^{\rm bkg}_{\rm exp}$ & UL [$10^{-6}$] \\
  \hline  
  $\Bp\to\Dm\ep\ep$   & 1.2 & 0 & $0.18\pm0.13$ & $<2.6$ \\
  $\Bp\to\Dm\ep\mup$  & 1.3 & 0 & $0.83\pm0.29$ & $<1.8$ \\
  $\Bp\to\Dm\mup\mup$ & 1.9 & 0 & $1.10\pm0.33$ & $<1.1$ \\
  \hline  
  \end{tabular}
  \caption{\belle\ results for the $\Bp\to\Dm\ellp\ellp$ search;
    $\epsilon$ is the signal reconstruction efficiency; $N_{\rm obs}$
    is the number of events in the signal region; $N_{\rm exp}^{\rm
    bkg}$ is the expected number of background events, and UL is the
    branching fraction 90\% CL upper limit~\cite{bib:belle}.}
  \label{tab:belle_result} 
\end{table}


\babar\ has performed the first measurement of the branching fraction
upper limits for the baryon- and lepton-number violating decays
$B^0\rightarrow\Lambda_c^+\ell^-$, $B^-\rightarrow\Lambda\ell^-$ and
$B^-\rightarrow\bar{\Lambda}\ell^-$~\cite{bib:babar2} with \nbb\
million \BB\ pairs. The $\Lambda_c^+$ and $\Lambda$ candidates are
reconstructed through the decay modes $\Lambda_c^+\rightarrow
pK^-\pi^+$ and $\Lambda \rightarrow p\pi^-$, respectively. The final
state tracks for both the $\Lambda_c^+$ and $\Lambda$ decays are
constrained to a common spatial vertex and their invariant mass is
constrained to the $\Lambda_c^+$ or $\Lambda$ mass. The candidates are
required to be within $\pm 15$~\mevcc of the nominal $\Lambda_c^+$
mass and $\pm 4$~\mevcc of the nominal $\Lambda$ mass. \Bmeson\
candidates are formed by combining baryon candidate with a $\mu^-$ or
$e^-$ and constraining them to a common point.  Bremsstrahlung energy
recovery is performed for electrons.  Background from
$e^+e^-\rightarrow e^+e^-\gamma$ events in the $\Lambda\ell$ channel
are eleminated by requiring more than four charged tracks in the
event.

Candidate selection is optimized using a figure of merit (see above)
with $a=5$. A neural net (NN) is used to provide further
discrimination between signal and background such that about 90\% of
the signal is retained and about 50\% of the background is
rejected. The remaining background after selection for the
$\Lambda_c^+\ell^-$ modes is composed of roughly equal amounts of \BB
and continuum events, while the background for the $\Lambda\ell$ modes
is almost entirely continuum.

The variables \DeltaE\ and \mes\ are used in the ML fit to the two
$\Lambda\ell$ modes; the $\Lambda_c^+\ell^-$ decay has more background
and the NN is used as a third discriminating variable. No significant
signal is observed and an upper limit is calculated for the branching
fraction for each decay mode, as shown in Table~\ref{tab:ul}.

\begin{table}[hbt]
\begin{center}
\begin{tabular}{lcccc}
\hline
Mode & $N_{\rm cand}$ & $\mathcal{B}$ ($\times 10^{-8}$) & $\epsilon$ (\%) & $\mathcal{B}_{90\%}$ ($\times 10^{-8}$) \\
\hline
$B^0 \rightarrow \Lambda_c^+ \mu^-$ & 814 & $-4_{-56}^{+71}$     & $26.3\pm0.9$ & $180$ \\
$B^0 \rightarrow \Lambda_c^+ e^-$   & 651 & $190_{-90}^{+130}$ & $25.7\pm0.7$ & $520$ \\
$B^-\rightarrow \Lambda \mu^-$       & 320 & $-2.3_{-2.5}^{+3.5}$ & $28.7\pm0.9$ & $6.2$ \\
$B^-\rightarrow \Lambda e^-$         & 194& $1.2_{-2.6}^{+3.7}$ & $27.2\pm0.6$ & $8.1$ \\
$B^-\rightarrow \bar{\Lambda} \mu^-$ & 192 & $1.5_{-1.7}^{+2.6}$ & $31.3\pm1.0$ & 6.1 \\
$B^-\rightarrow \bar{\Lambda} e^-$   & 74  & $-0.9_{-0.0}^{+0.7}$ & $30.0\pm0.6$ & $3.2$ \\
\hline
\end{tabular}
\caption{Summary of number of candidates ($N_{\rm cand}$), branching
fraction central value ($\mathcal{B}$), signal efficiency
($\epsilon$), and branching fraction 90\% CL UL
($\mathcal{B}_{90\%}$)~\cite{bib:babar2}.}
\label{tab:ul}
\end{center}
\end{table}


\begin{figure}[htb]
\begin{center}
    \includegraphics[width=0.62\textwidth]{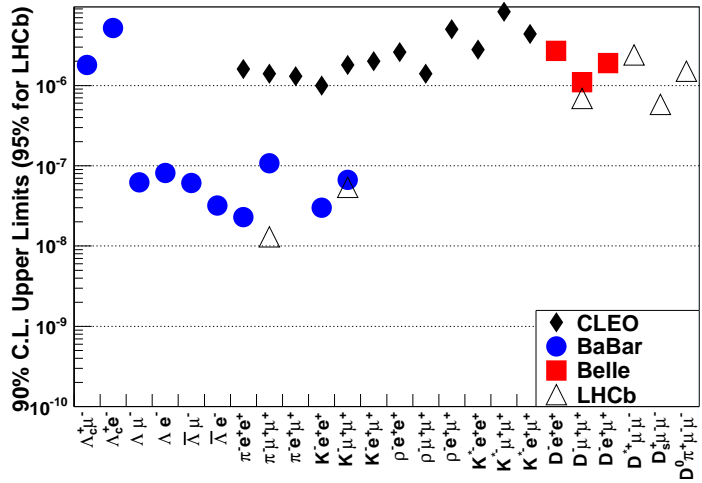}
  \caption{Branching fraction upper limits for \Bmeson\ LNV decays
  from CLEO~\cite{bib:cleo}, \babar~\cite{bib:babar1,bib:babar2}, \belle~\cite{bib:belle}, and LHCb~\cite{bib:lhcb}.}
  \label{fig:summary}
\end{center}
\end{figure}


\end{document}